\documentclass[conference]{IEEEtran}
\IEEEoverridecommandlockouts
\usepackage{cite}
\usepackage{amsmath,amssymb,amsfonts}
\usepackage{algorithmic}
\usepackage{graphicx}
\usepackage{textcomp}
\usepackage{xcolor}
\PassOptionsToPackage{hyphens}{url}\usepackage{hyperref}
\def\BibTeX{{\rm B\kern-.05em{\sc i\kern-.025em b}\kern-.08em
    T\kern-.1667em\lower.7ex\hbox{E}\kern-.125emX}}
\begin{document}

\title{
Emotion-Guided Music Accompaniment Generation Based on Variational Autoencoder
}
\author{\IEEEauthorblockN{Qi Wang, Shubing Zhang , Li Zhou \IEEEauthorrefmark{1}}
\IEEEauthorblockA{
\textit{China University of Geosciences(Wuhan)} \\
\{wangqi233,zhouli\}@cug.edu.cn}

\thanks{\textsuperscript{*} Corresponding author}

\thanks{\textsuperscript{1} This research was funded by the Chinese Regular Projects of the Humanities and Social Sciences Fund of the Ministry of Education of Grant No.16YJAZH080.}
}

\maketitle
\begin{abstract}
Music accompaniment generation is a crucial aspect in the composition process. Deep neural networks have made significant strides in this field, but it remains a challenge for AI to effectively incorporate human emotions to create beautiful accompaniments. Existing models struggle to effectively characterize human emotions within neural network models while composing music. To address this issue, we propose the use of an easy-to-represent emotion flow model, the Valence/Arousal Curve, which allows for the compatibility of emotional information within the model through data transformation and enhances interpretability of emotional factors by utilizing a Variational Autoencoder as the model structure. Further, we used relative self-attention to maintain the structure of the music at music phrase level and to generate a richer accompaniment when combined with the rules of music theory.
Our experimental results indicate that the emotional flow of the music generated by our model has a strong correlation with the input emotion, demonstrating the model's strong interpretability and control of emotional flow. The generated music is also well-structured, diverse, and dynamic, outperforming the baseline models.

\end{abstract}

\begin{IEEEkeywords}
Music Accompaniment Generation, Emotional Flow, Variational Autoencoder, Rule constraints
\end{IEEEkeywords}

\section{Introduction}
Music evokes emotions in listeners, making it a powerful and intuitive medium for understanding. It also serves as a driving force for musicians to create. One important aspect of composing is incorporating emotional expression into the music. Composers use their emotions along with their technical skills and knowledge to craft their compositions. 

Current AI methods fall short of replicating a composer's approach. Neural networks primarily focus on combining and utilizing pre-existing knowledge of compositions, rather than incorporating emotions as high-level information. Our research aims to overcome this limitation by developing a model for generating accompaniment that takes emotions into account.

The way emotions are processed impacts every aspect of music composition and, as a result, every aspect of deep neural networks \cite{b3,b4,b5,b6}. This puts a significant emphasis on the need for network control. While autoregressive models can effectively capture key elements of music, they lack transparency and do not guarantee internal control and interpretability of musical information. Adversarial networks \cite{b7,b8} can separate elements like pitch, rhythm, and texture, but they struggle with capturing emotional information and prioritize interpretability over musicality and structure.

Additionally, many music generation models \cite{b2,b3,b4,b5,b6,b7,b8,b9,b10,b11,b12,b13,b14,b15} primarily focus on identifying and evaluating the emotional aspects of music, rather than using them as a controllable variable. Therefore,
instead of using subjective and limited emotional labels\cite{b1}, such as "relaxed" or "nervous," we have adopted Thayer's continuous emotion model\cite{b2}. This model takes into account two quantitative and controllable factors: valence, which measures the level of positivity or negativity, and arousal, which measures the level of excitement or calmness. This approach provides a controlled understanding of human emotions.

Thus, we designed a system based on Variational Autoencoder, a controllable deep learning model, which incorporates emotional factors into the neural network's learning process. The user inputs valence and arousal trends, which are then encoded using our Valence Encoder and Arousal Encoder. The model then decodes and reconstructs this information to generate 2-bar piano accompaniments that match the emotional flow of the user's input.
To compose a dynamic piece of music, we take into account two key elements: tonality\cite{b9}, which enhances the beat and rhythm of the music by incorporating rule-based constraints in the model's decoder, and structural organization\cite{b10}, which improves the storytelling aspect of the music and preserves the internal structure of the piece through a self-attention mechanism.
Our data, code, and samples have been made publicly available \footnote[2]{\url{https://github.com/Duoluoluos/Emotion-Guided-Music-Accompaniment-Generation}}online.

Our main contributions include:
\begin{itemize}
    \item Emotion-Guided Composition, where the user inputs an Emotion-Flow Curve and the model generates music
that closely matches the input emotions.

\item 	Enhanced accompaniment generation, incorporating global tonality, music phrases, and local texture for a more realistic and dynamic improvised accompaniment.

\item Integration of rules and deep learning, combining the creative capabilities of deep networks with the constraints of music theory to improve the transparency of the music creation process.
\end{itemize}
\section{Related Works}

\subsection{Accompaniment Generation}

Generating musical accompaniment is essentially a specific type of music generation problem\cite{b26}, where the melody is used as a constraint, and the accompaniment is the generated music. In the past, accompaniment generation was approached in the same way as music generation, treating pitch and temporal values as simple data. Algorithms such as Hidden Markov Chain (HMC)\cite{b28}, Random Forest (RF), Support Vector Machine (SVM)\cite{b27} \cite{b29}, etc. were used to approach the problem from a regression perspective. However, with the advancement of deep learning, more accurate prediction models have been developed.

DeepBach\cite{b11}, a well-known music generation network based on RNN/LSTM\cite{b3,b4,b5,b6} networks, represents Bach choral as voice lists with metadata lists and embedding representation to RNN for prediction. However, RNN/LSTM networks alone may not be sufficient for achieving the required level of long-range coherence in accompaniment. Hybrid models, such as the RNN-LSTM model in paper \cite{b5} and the RNN-RBM model in paper \cite{b30}, have been proposed to address this issue. The RNN-LSTM model learns different models in stages, while the RNN-RBM model uses several Restricted Boltman Machines (RBMs) and samples the output of the RBMs as input for the RNN, training local information and then makes autoregression for each information.

In 2018, the Music Transformer \cite{b12} was introduced, which shifted the focus from regression problems and note prediction to natural language processing (NLP) techniques for recognizing relationships between different segments of music and evaluating the logicality of musical phrases, similar to how NLP tasks analyze relationships and coherence in language. The Transformer model uses attention mechanisms, positional coding, and other techniques to ensure long-range coherence, making it useful for various accompaniment generation tasks such as drum and piano accompaniment. The model is similar to text completion in NLP, using a priori melodic data and key information such as drum beats to "fill in" missing features. Papers \cite{b13,b14,b15} have expanded upon this data representation and the MuMidi proposed in paper \cite{b33} can solve harmonic problems in a long-term context by integrating pitch, time value, and tempo. However, the generation process is not always interpretable or controllable and the randomness of notes can increase over time, resulting in non-sequential music.

To improve control over the music generation process, various methods have been employed. MuseBert \cite{b16} uses data corruption and fine-tuning during the inference learning process, while Music VAE \cite{b8} \cite{b17,b18,b19,b20} uses decoupled feature representations such as pitch, chord, and texture, and employs interpolation, back-and-forth sampling, and temperature factors to increase accompaniment diversity. MuseGAN \cite{b21} treats music data as images and can generate multi-track accompaniments, but the structure of each track is not well-constrained by composition rules and the resulting music may not be as listenable. It is worth noting that the "hidden space" of the Variational Autoencoder(VAE) is better suited to the music generation problem than the image representation method used in the generative adversarial network. Unlike pass-through data, notes are affected by pitch, time, and velocity and have a high dimensionality of information. The VAE \cite{b17} normalizes this information to the hidden space for posterior estimation and reconstruction using an Encoder-Decoder architecture, which can be combined with a "learning from scratch" strategy and improve the model's ability to migrate and transfer. Therefore, we chose to use VAE as a controllable accompaniment generation model. Our model can generate well-structured accompaniments that conform to certain composition rules and follow an Emotion Flow.

\subsection{Emotional Flow Guided Composition}
Valence and Arousal are commonly used as quantitative measures of musical emotion in research. Studies\cite{b34} have shown that the rhythmic density of music, determined by the duration of notes in each measure, can affect a person's arousal levels independently of note velocity. Additionally, the melodic and harmonic direction of a song can affect the overall emotional direction \cite{b35}, referred to as valence. These factors can have a significant impact on the emotional response to a piece of music.

The objective of our research is to extract features from Emotion Flow, specifically the Valence Curve and Arousal Curve \cite{b24,b25}, and then systematically associate those features with the generated accompaniment. Previous research, as shown in the paper \cite{b22}, used dynamic programming and template-matching methods to complete the Emotion-Flow Guided Accompaniment Generation. However, these methods can ensure the audibility of the music but do not guarantee the diversity of the accompaniment. In contrast, deep neural networks can achieve accompaniment diversity through large-scale learning, but they struggle to maintain the structure of the music compared to methods such as template matching \cite{b26}. Although self-similarity \cite{b8} can maintain some of the structure, neural network methods have difficulty ensuring the structure of the music because the music structure is strongly regulated through music phrases. Therefore, decoding music segments into "phrase" units is the key to maintain music structure. In this paper, we propose using a VAE which makes full use of structured features of the music to improve the overall structure and diversity of the accompaniment.

\section{Methods}

\subsection{Data Preparation} \label{sec3.1}
The POP909 Dataset \cite{b35} comprises 909 popular music tracks, which are piano-based and have a total running time of 60 hours. Each track is stored in MIDI file format and includes three separate components - melody, bridge, and piano. The bridge and piano tracks serve as an accompaniment. Additionally, the dataset includes chord and bar annotations for each song.

The POP909 dataset includes melodies that are broken down into 2-bar, 4-bar, and 6-bar fragments. The bar annotations in the dataset provide information about the structure of these fragments. The chord annotations, on the other hand, provide information about the harmony of each bar in the melodies.

To address the issue of music structure in a consistent manner, we discovered that the majority of music is composed of 2-bar segments. As a result, we carried out data cleaning, filtering out 2/4-bar segments and 2/4-bar segments with 6-bar introductory fragments. The training and testing sets were then split in an 8:2 ratio.

As sample data, we selected a subset from the Nottingham Dataset \cite{b36}. This dataset comprises over 1000 European and American folk songs, all of which have chord annotations. For validation purposes, we chose 2-bar and 4-bar segments from the dataset. The collated data information is presented in Table \ref{The Specific information of Selected Dataset}. (It is worth noting that if the user-supplied music does not have chord annotations like the sample data, we used Bi-LSTM Harmonizer \cite{b42} to implement the chord annotations)

\begin{table}[htbp]
\caption{The Specific information of Selected Dataset}
\label{The Specific information of Selected Dataset}
\begin{center}
\begin{tabular}{cccc}
\hline
Usage                 & Source             & Quantity & Duration(Mean) \\ \hline
Training                 & POP909 Dataset     & 685      & 225s           \\
Validation            & POP909 Dataset     & 172      & 201s           \\
Samples & Nottingham Dataset & 20       & 72s            \\ \hline
\end{tabular}
\end{center}
\end{table}    

To showcase the capabilities of our model, we chose two representative songs, one with high valence and the other with low valence, from the 20 songs we used. These songs were made available on a web page for users to evaluate and \footnote[3]{\url{https://soundcloud.com/ko9isjyplxrb/sets/demos-of-emotion-guided-generated-accompaniment}}enjoy.

\subsection{Models}

\begin{figure*}[hbtp]
\centerline{\includegraphics[scale=0.65]{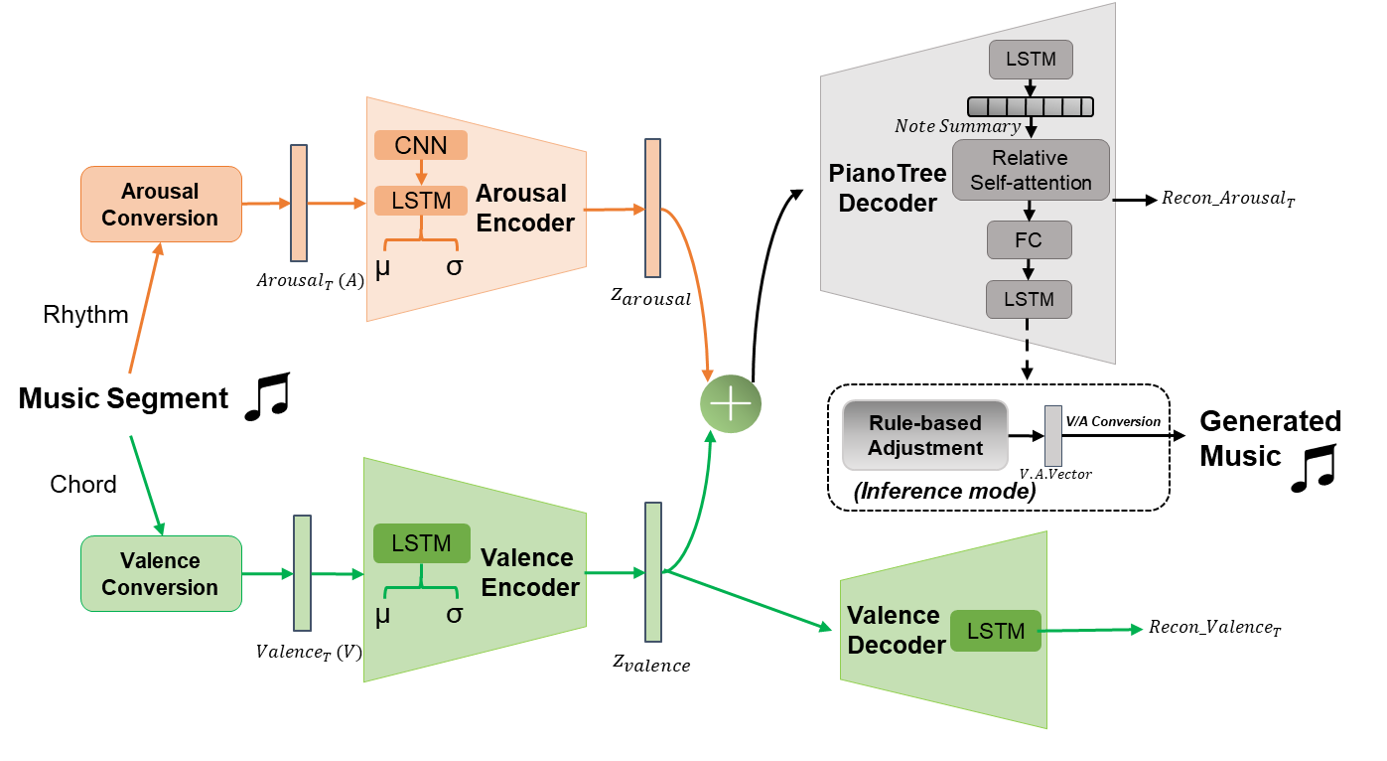}}
\caption{Overall Architecture. The model begins by preprocessing the data with a converter to obtain valence and arousal values. These values are then encoded into a hidden space by their respective encoders and subsequently reconstructed by the decoder after being fused together. To improve the overall listening experience, we adjust the tonality dynamically using music theory-based adjustment.}
\label{Overall Architecture}
\end{figure*}

\subsubsection{The Conversion of Valence and Arousal}
The overall architecture is illustrated in Figure \ref{Overall Architecture}.  
The initial music data is represented by piano rolls. Each row of the piano roll matrix corresponds to one of the 128 pitch values and each column corresponds to a unit of time, with the duration of a 16th note used as the unit of time. The accompaniment tracks were merged and transformed to produce the accompaniment piano roll $p_{T}^{ACC}$, where T represents the duration of the altered accompaniment fragment. Similarly, the rhythm piano roll is represented as $p_{T}^{RHY}$, and the labeled chord progression is represented as $c_{T}$. According to the twelve-mean meter \cite{b36}, $c_T$ is a matrix of 12 $\times$ T, where 12 is the number of notes in an octave.
\begin{equation}
    Valence_{T}=V(\bar{c_T})
\end{equation}

Where $V(\cdot)$ is the Valence's mapping and  $\bar{c_T}$ is the chord data after normalizing the root note of $c_T$ to the C3 note. This is to ensure that the Valence is in the same key, and we set the T here to 8.

Also with respect to Arousal's mapping as $A( \cdot )$, there are,

\begin{equation}
    Arousal_{T}= A(p_{T}^{ACC}+p_{T}^{RHY})
\end{equation}

The operation of mapping A is to transform the multitrack music data into a tree structure \cite{b37}, where the nodes of the tree can more clearly characterize the density distribution of notes. Arousal is a four-dimensional matrix of size $128\times T \times 16 \times 8$, denoting the pitch-duration-density grouping, respectively.

Denote the quantization operation of Arousal and Density as $| \cdot |$, 

\begin{equation}
    |Arousal|_T=\frac{1}{5 \cdot T}  \sum_{T} \sum_{pitch} A(p_{T}^{ACC}+p_{T}^{RHY})
\end{equation}
\begin{equation}
    |Valence|_T = \sum_{T} W_{V(\bar{c_T})}
\end{equation}

The $W$ value in this context refers to the chroma weights of each chord and serves as a measure of the valence, or emotional assessment, of each chord. By performing a quantization-transformation operation, the emotional content of the music can be translated into a format that the composition model can understand, allowing for the user's desired Emotion Flow to be incorporated into the final output.

\subsubsection{Valence/Arousal Encoder}

Arousal and Valence Encoder are both dominated by LSTM as the backbone network. Arousal Encoder extracts the features of pitch-time-value information through a CNN with a (4,12) sized kernel  in convolutional layer and (1,4) sized kernel in max pooling layer.

In fact, after the features are extracted by the convolutional network, the arousal information is more concise and refined [38], so that Decoder can learn better emotional features.

The layers of the LSTM network are all 1, and both are bidirectional. the dimension of the input weight of the Arousal Encoder is 256, and the dimension of the output weight is 1024. the dimension of the input weight of the Valence Encoder is 32, and the dimension of the output weight is 1024. Both are encoded to calculate the mean and variance of the probability distribution and sampled to obtain a 256-dimensional latent space variable $z_{Arousal}$ or $z_{Valence}$.

\subsubsection{Decoder}

The Valence Decoder is introduced first, and the LSTM encoder of the decoder is roughly the same, except that the input side is fused with $z_{valence}$, and the dimension is modified to 292. The reconstructed Valence is estimated by calculating the variance and mean, and it is input to the LSTM as a token so that the decoding part of the model is completed. The probability distribution of valence is a 12-dimensional Bernoulli distribution.

PianoTree Decoder, on the other hand, refers to the design of the paper \cite{b39} and uses the model in this paper as a baseline. The original model is divided into two main stages, one is the time domain decoding and the other is the decoding of notes for each pitch. Since different notes may be concatenated into fragments and have some autocorrelation in the structure of the music to form the music phrases, we performed a note summary operation after the time-domain decoding operation and introduced a self-attention mechanism, which we will explain the ins and outs in detail in the next subsection.

The role of the first Pianotree-LSTM in Figure \ref{Overall Architecture} is to decode 512-dimensional latent space vectors. latent space vectors are the hidden space mapping changes of notes, and LSTM (hidden size=1024) is to summarize and summarize the results of the changes in the temporal dimension, so we call the summarized results note summary with size (1,512). After obtaining the relative self-attention, it is then decoded in the dimension of the pitch by LSTM(2) and mapped to 128 pitches through the fully connected layer. For each or each class of notes, respective temporal values are then decoded by LSTM (hidden size=16) to obtain the emotional stream/music sequence after reconstruction.

\subsubsection{Relative Self-Attention}
In order to maintain the structural organization in the music sequences, we introduce a self-attentive mechanism. This inspiration comes from the paper \cite{b9}, which does this by comparing a template music sequence fragment with a training music fragment and obtaining the correlation of the relative positions in the two sequences by one-dimensional/two-dimensional convolution, and the resulting correlation data is called self-similarity.

In this paper, self-similarity is not done by convolution operation because we do not have template fragments, but by note summary, a tensor of stacked pitch and mood information in the time domain. Similarly, since self-attention obtains the autocorrelation information inside the input by soft addressing, it is just possible to obtain the autocorrelation of note summary in the time domain and thus maintain the structured organization of the music fragments as the estimated "music phrases".
 
Since there is some time invariance in the relative positions of the sequences \cite{b12}, we also introduce offsets. Each fragment is not very informative, and to optimize the efficiency of the algorithm, we  use a single-head attention mechanism. The query, key, and value tensor of relative attention are written as Q, K, and V, respectively. $S^{rel}$ represents the offset matrix and the matrix element $r=NS_{k}-NS_{q}$, where $NS_k$ and $NS_q$ are the note summary query and key's position code, then the formula for relative self-attention(abbreviated as $Att$) is

\begin{equation}
    Att = Softmax(\frac{QK^T+S^{rel}}{\sqrt{D}})V.
\end{equation}

As for the parameter settings, we set the weight dimension of Q to 1024 and the weight dimension of K, V to D=128.

\subsubsection{Rules-based Constraint}
Two rules are very common in the realm of improvised accompaniment, enriching the player's accompaniment performance by changing tonality. The first principle is to add variety to the chords by making small adjustments to the chord tuning. 
The second technique is to add a sense of layering between the different voices by shifting the tonality of the chords significantly at the same time. 

Either way, chord arrangement is the most important thing.
If we want to use the rules in our accompaniment generator, we need to grasp the key information and build the model. Whether it's chord transposition or pitch shifting, it's essentially shifting pitch. So instead of inferring from the model, we can use the chord arrangement and transposition information directly to shift the pitch and change the generated accompaniment.

To obtain the chord transposition information, a mathematical evaluation is required. We note that the originally labeled chords of the input melody are $C^{pre}$ and the chords generated by PianoTree decoding are $C^{gene}$, each chord is represented by 12 mean meters, so it is a 12-dimensional vector. The two are compared and the maximum difference is used as the criterion for transposition. Note the current bar number $i$, the pitch shift $\Delta C$ refers to:

\begin{equation}
    \Delta C = argmax(\frac{C^{pre}_i C^{gene(T)}_i }{|| C^{pre}_i || \cdot || C^{gene(T)}_i ||})
\end{equation}
	 Here T denotes matrix transposition.
Each bar has a chord best transposition selection, and a number of bars with large $\Delta C$ are selected for pitch shift so that tonality adjustment is achieved by means of rules and mathematical modeling.

\subsection{Training Objective}
The training objective of VAE \cite{b40} is much the same, and its loss function mainly consists of regularization loss and reconstruction loss. To shorten the formulation, we abbreviate Valence and Arousal as V and A.

For the regularized loss, we set the prior Gaussian distributions of Valence and Texture as $p(z_{V})$ and $p(z_{A})$, and the posterior distributions after encoder are noted as $p(z_V|V)$, $p(z_{A}|A)$, respectively. To find the regularization loss of the two probability distributions, we commonly use the KL scatter [40], denoted here as KL($\cdot$).
 
For the reconstruction loss, we set the probability distribution of the Valence Decoder output as $p(V|z_V)$ and the PianoTree Decoder output as $p(A|z_A, z_V)$, and the reconstruction loss is generally found by finding the log probability expectation value. In summary, the loss function $Loss(V, A)$ of the model is

\begin{equation}
\begin{split}
Loss(V, A) = E_p[log p(V|z_V) + log p(A|z_V,z_A)]\\
+ KL(p(z_V|V) || p(z_V)) + KL(p(z_A|A) || p(z_A))  
\end{split}  
\end{equation}

\section{Experiments}
\subsection{Training Details of Our Proposed Model}
The experiment was run on a host with a 12th Gen Intel(R) Core(TM) i7-12700H and a single NVIDIA GeForce RTX3060 6GB.
 
In the section \ref{sec3.1}, we explain the dataset and convert the MIDI files in the dataset into a piano roll representation and a 12-measure chord representation, respectively We set the batch size to 128, so that the model is trained with a time value of 32 for each arousal fragment and 8 for the valence fragment.

When training our VAE model, we set the epoch to 6 and the learning rate to $10^{-3}$ with an exponential decay of 0.999 and a minimum value of $10^{-5}$. To speed up the training speed and reduce the possibility of model divergence, we use the Teacher-Forcing strategy. The Teacher-Forcing training ratio of Encoder-PianoTree Decoder , and Encoder-Valence is set to 0.6 and 0.5 respectively. The training ratio of Encoder-Valence Decoder is set to 0.5.

\subsection{Baseline Models}

Our baseline models are Poly-dis and M-GPT chosen from the model in the paper \cite{b41} \cite{b12}. Poly-dis, the state-of-the-art disentanglement learning-based model, decouples the characterization of harmony and texture. Unlike our rule constraint and modeling, this model achieves the adjustment of the generated accompaniment by learning prior and posterior sampling. M-GPT is the state-of-the-art piano music generation model and can harmonize the melody using auto-regression principles. 

\subsection{Emotional Flow Comparison Test}

The experiment aims to compare the correlation between the Emotional Flow entered by the user, used as a guide, and the Emotional Flow finally generated by the system. This is an important indicator of the effectiveness of the system's control over the input Emotional Factors.

We evaluate the correlation by comparing the Pearson coefficients between the two sequences, referring to the evaluation metrics in the paper \cite{b22}, so as to avoid misevaluation due to misalignment of the Emotional Flow.

There are two constraints on the Emotional Flow of the user input guidelines. The first is that there cannot be more than five extreme points per flow curve, except for the start and end points. This is because the melodic data of the sample data does not exceed 90s in length, and too many extreme  points mean too many melodic ups and downs, which is not in accordance with the rules of music composition. The second is that each flow curve must have a certain amount of ebb and flow, because too much flatness is not necessary for correlation. Specifically, $\bar{V}$ and $\bar{A}$ are the mean values of the valence and arousal curves, and the duration of the melody is set to T.
\begin{equation}
	\begin{cases}
	\frac{1}{T} \int_{0}^{T} (V-\bar{V})^2 dt > 0.15\\
	\frac{1}{T} \int_{0}^{T} (A-\bar{A})^2 dt > 0.15
	\end{cases}
\end{equation}

The data for the experiment were obtained from the "Samples" mentioned in the section \ref{sec3.1}, with 20 pieces of music to be validated. Four typical cases were selected to visualize the results. The criteria we chose are similar to the idea of control variables, which are the correlation of Arousal Flow in the low arousal and high arousal cases, and the correlation of Valence Flow in the Low Valence and High Valence cases, respectively. We calculated the average valence and arousal correlation values for 20 samples of music. For statistical convenience, high arousal/valence is denoted as High Input Basis (HIB) and low arousal/valence is denoted as Low Input Basis (LIB).

\begin{figure}[htbp]
\centerline{\includegraphics[scale=0.55]{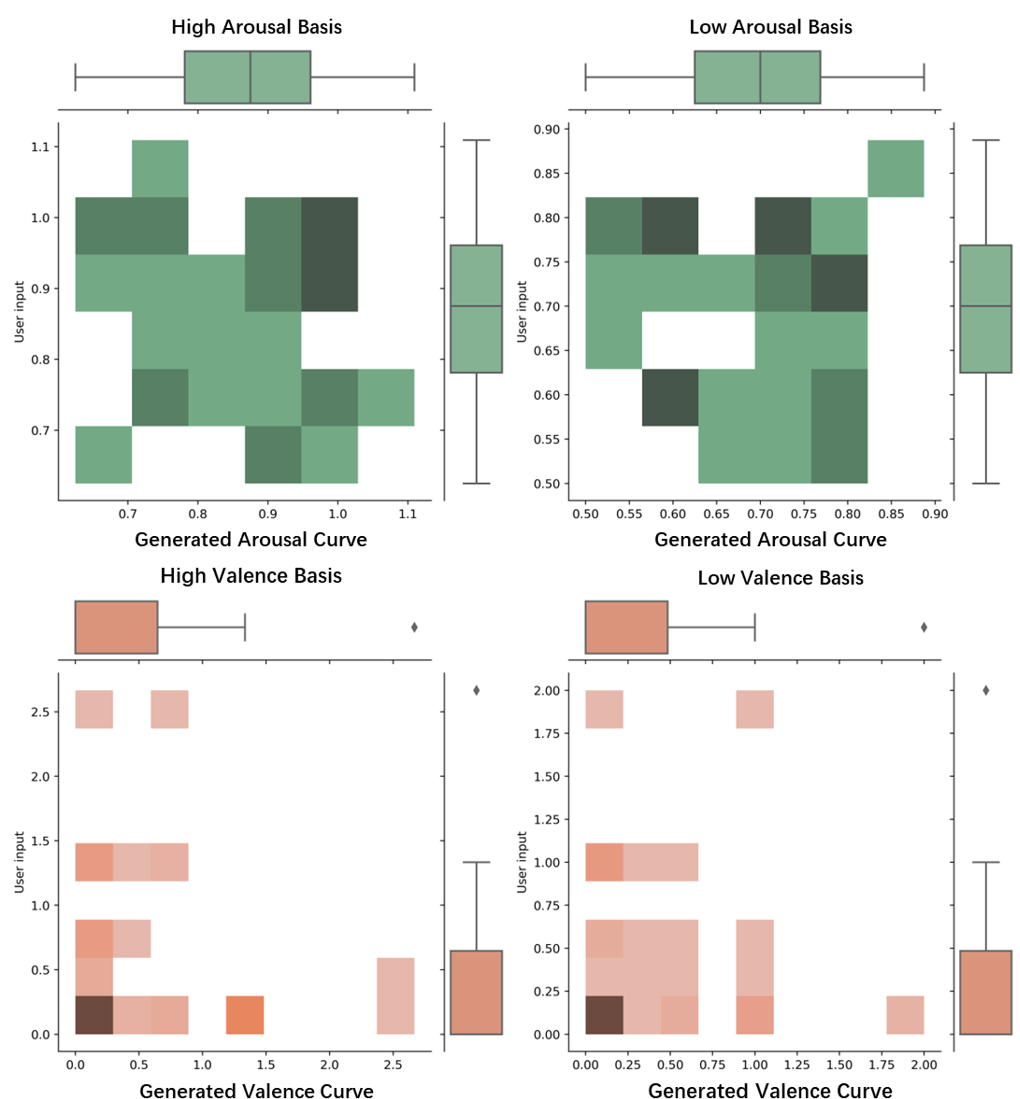}}
\caption{Visualization of Emotion Flow Comparision}
\label{Visualization of Emotion Flow Comparision}
\end{figure}

The visualization in Figure \ref{Visualization of Emotion Flow Comparision}, a combination of a heat map and box plot, presents a comparison of the input and output Emotional Flow. The heat map illustrates the specifics of the Emotional Flow, while the box plot offers a broader statistical comparison. The results reveal that the mean values and quartiles of the Emotional Flow are similar for both the user input and the system output. This suggests that the system-generated Emotional Flow aligns with the user input statistically, regardless of the Emotional Flow's baseline.

We also compared the association values between the baseline model and our VAE model, as shown in Table \ref{Emotional Flow correlation coefficients between input and output for different scenarios}. Where the baseline model is abbreviated as Poly-Dis, our model is called VA-VAE.

\begin{table}[htbp]
\caption{Emotional Flow correlation coefficients between input and output for different scenarios}
\label{Emotional Flow correlation coefficients between input and output for different scenarios}
\centering
\begin{tabular}{c|cccc}
\hline
\textbf{Emotion Type}  & \textbf{Model} & \textbf{HIB}   & \textbf{LIB}   & \textbf{Mean}  \\ \hline
Valence & Poly-Dis   & 0.764  & 0.619   & 0.688   \\  
    & M-GPT  & 0.512 & 0.624 & 0.442           \\
    & VA-VAE & 0.682 & \textbf{0.804} & \textbf{0.715} \\   \hline
Arousal & Poly-Dis   & 0.571  & 0.647   & 0.606     \\  
  & M-GPT  & 0.364 & 0.577 & 0.591  \\
& VA-VAE   & \textbf{0.627} & 0.631   & \textbf{0.628} \\ \hline
\end{tabular}
\end{table}

It can be seen that the average correlation of our model outperforms the baseline models for both valence flow and arousal flow. The correlation of our VA-VAE also outperforms the baseline model under HIB versus LIB.

\subsection{Subjective Musicality test}
The subjective musicality assessment was mainly a professional assessment by music experts. A total of 44 junior and senior music majors and graduate students were invited. The music experts were randomly selected from two of the eight sample groups, and each group contained two pieces of music, one with the accompaniment generated by the baseline Transformer model and the other with the accompaniment generated by the VA-VAE model. The two pieces of music were not distinguished by name; in other words, the music experts' music was selected in a completely blind manner.
The music experts evaluated the level of the accompaniment from four angles: 1) whether the overall layout of the composition was appropriate; 2) whether the chords were harmoniously chosen and connected; 3) whether the rhythmic density (articulation points) was specific to the melody; and 4) whether there was a sub-melody or passing phrase that accentuated the melody. Each evaluation angle is evaluated quantitatively using a rating value, and is assigned a score of 1 to 5. The above four perspectives are abbreviated as Q1, Q2, Q3 and Q4.

\begin{figure*}[htbp]
\centering
\centerline{\includegraphics[scale=0.8]{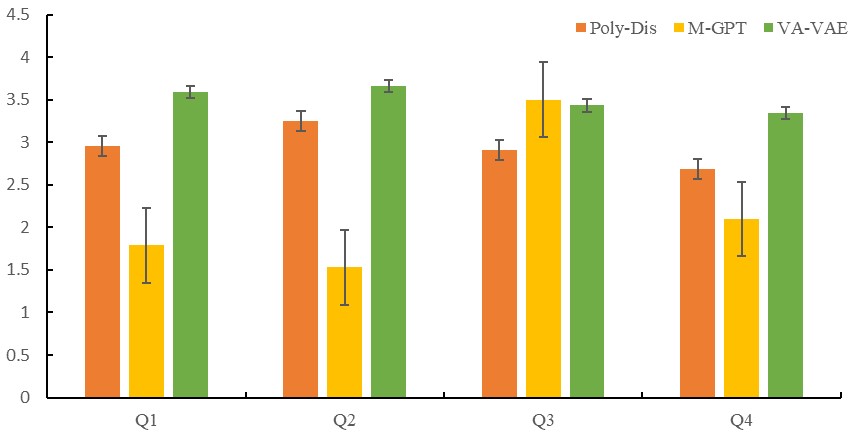}}
\caption{Subjective Musicality Analysis by Music Majors}
\label{fig:Subjective Res}
\end{figure*}

The experimental results are shown below, and the final score for each assessment perspective is based on the weighted average score.

From the experimental results shown in Fig \ref{fig:Subjective Res}, we can see that the weighted average score of our VA-VAE model is stronger than that of the Baseline models in terms of the overall layout of the weave (Q1), chord selection and connection (Q2), melodic counterpoint (Q3), and melodic underscoring (Q4). The overall arrangement of the accompaniment generated by our model is more reasonable, and the chord selection and connection are more fully considered, and the rhythm between the accompaniment and the melody is more organized and regular, which can also better support the melody. The musical accompaniment generated by our model has a more artistic character.

\begin{figure}[htbp]
\centering
\centerline{\includegraphics[scale=0.6]{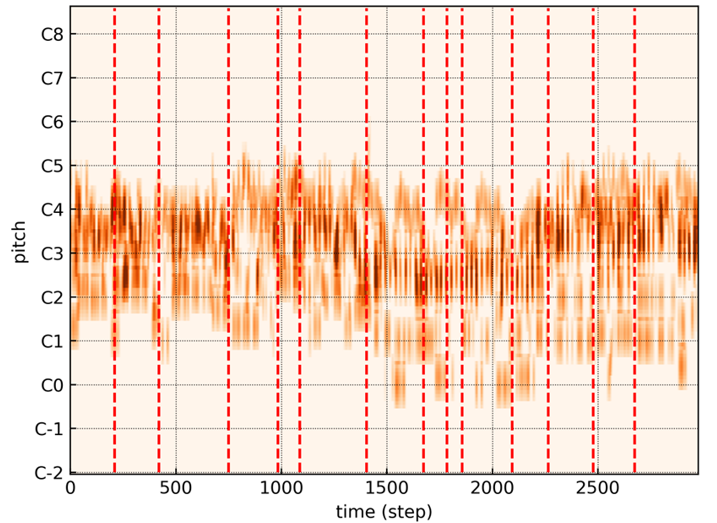}}
\caption{Visualization of relative self-attention}
\label{Visualization of relative self-attention}
\end{figure}

Refer to Figure \ref{Visualization of relative self-attention} for a visual representation of the music's attention structure.

The darker the color of the music phrases, the greater the weight of attention. The structure of the different "music phrases" gathered by attention mechanism is divided by dotted lines, so that the music as a whole is well organized.

\subsection{Ablation Study}

For the ablation study, we abbreviated the control group without relative self-attention and Rule Constraint (RC) as CG, the model after adding relative self-attention as CG+NS, and then after adding Rule Constraint as CG+NSR. We used a quantitative approach to assess the generation The quality of the accompaniment in the ablation experiment is assessed quantitatively. Quantitative metrics such as pass/fail ratios, null ratios, etc. are less applicable in our piano improvisation accompaniment generation task. The key criteria for the evaluation of the accompaniment task are the texture of the accompaniment, the harmony of the accompaniment with the melody, the contribution to the melody, etc. This way of evaluation is very similar to that of the translation task, where the harmony of the accompaniment is like the valuation of the translated utterance, the weaving arrangement is like the wording of the translation, and the contribution to the melody is like the synthesis and comparison of the information in the translation task. Therefore, we chose the MUTE evaluation index from the paper \cite{b43}, which is analogous to the F-Score evaluation index in the translation task, to accurately and quantitatively assess the level of the accompaniment arrangement.

In MUTE, F1 Score(FS) evaluates the "translation accuracy" of the accompaniment from the perspective of 128 pitches and is suitable for evaluating texture, while the F1 Score Pitch Class(FSPC) normalizes the pitches to 12 basic pitches and is therefore suitable for evaluating harmony.

\begin{table}[htbp]
\caption{Results of Ablation Study}
\label{Results of Ablation Study}
\centering
\begin{tabular}{c|ccc}
\hline
\textbf{Emotion Type}  & \textbf{Model} & \textbf{FS}   & \textbf{FSPC}    \\ \hline
High Valence & CG  & 0.2265    & 0.2656   \\  
        & CG+NS   & 0.1865   & 0.3521  \\ 
        & CG+NSR   & 0.2422   & 0.3542  \\ 
        \hline
Low Valence & CG  & 0.175   & 0.2453  \\ 
& CG+NS    & 0.1865   & 0.3521       \\ 
& CG+NSR    & 0.2078   & 0.3672       \\ 
\hline
\end{tabular}
\end{table}

\begin{figure}[htbp]
\centering
\centerline{\includegraphics[scale=0.3]{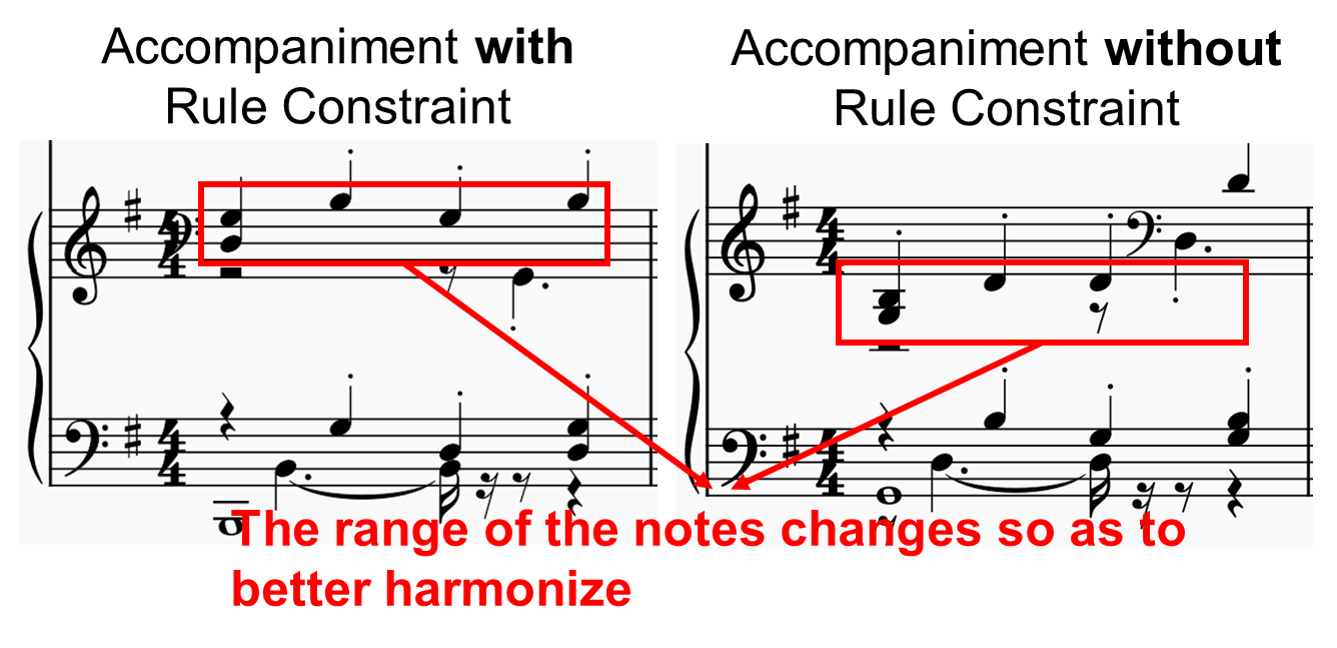}}
\caption{Visualization of Ablation Test on RC }
\label{Visualization of Ablation Test on RC }
\end{figure}

As seen in Table \ref{Results of Ablation Study}, the model incorporating relative self-attention and RC outperformed the CG and CG+NS control groups in both FS and FSPC metrics. Whether it is harmony or texture, the newly incorporated relative self-attention mechanism and rule constraint can be better designed and orchestrated to create higher quality accompaniment. Further, we visualized the comparison test of the rule constraints, as shown in Figure \ref{Visualization of Ablation Test on RC }, and found that the rule constraints did indeed shift the range of the accompaniment to better harmonize the melody.

\section{Conclusion}
In this study, we investigate the generation of musical accompaniment that is guided by emotional flow. We focus on two key aspects of the problem. First, we establish a mechanism for converting emotional streams into music information data and a VAE network architecture that is tailored to emotional quantization data, allowing us to control the network model with emotional factors. Secondly, we optimize the structural planning of accompaniment generation by introducing the Self-Similarity and relative self-attention mechanism. By using rule constraints, we further improve the local and global tonality of the music. This approach of progressing from the whole to the local, layer by layer, allows us to create an automatic accompaniment system that has excellent emotional flow control and high-quality music generation.
 
In the future, we plan to further improve our research. Currently, the accompaniment is generated by a single instrument and we intend to extend it to include multiple instruments to create an automated orchestra. Additionally, the representation of emotional flow is not yet clear, and we will research on better visualization methods to make the AI technology more user-friendly.

\section*{Acknowledgment}
This research was funded by the Regular Projects of the Humanities and Social Sciences Fund of the Ministry of Education of Grant No.16YJAZH080.

\end{document}